\documentstyle[rotate,graphicx]{mn}

\def\msun{{\rm M_{\odot}}}

\title[The Brightest Black Holes]
{The Brightest Black Holes}
\author[A.R.~King]{
A.R.~King\\
Theoretical Astrophysics Group, University of Leicester,
Leicester, LE1~7RH, UK}

\begin{document}

\maketitle

\begin{abstract}
I suggest that there are two classes of ultraluminous X--ray sources
(ULXs), corresponding to super--Eddington mass inflow in two
situations: (a) thermal--timescale mass transfer in high--mass X--ray
binaries, and (b) long--lasting transient outbursts in low--mass
X--ray binaries. These two classes are exemplified by SS433 and
microquasars like GRS~1915+105 respectively. The observed ULX
population is a varying mixture of the two, depending on the star
formation history of the host galaxy. ULXs in galaxies with vigorous
star formation (such as the Antennae) are generally SS433--like, while
ULXs in elliptical galaxies must be of the microquasar type. The
latter probably have significantly anisotropic radiation
patterns. They should also be variable, but demonstrating this may
require observations over decades. The close analogy between models of
X--ray binaries and active galactic nuclei (AGN) suggests that there
should exist an apparently super--Eddington class of the latter, which
may be the ultrasoft AGN, and a set of X--ray binaries with
Doppler--boosted X--ray emission. These are presumably a subset of the
ULXs, but remain as yet unidentified.
\end{abstract}
\begin{keywords}
accretion, accretion discs -- binaries: close -- galaxies, active --
X--rays: stars -- X--rays: galaxies
\end{keywords}

\section{Introduction}

Letting matter fall on to a black hole is the most efficient way of
converting its rest--mass energy into radiation. Accretion is thus
assumed to power the most luminous objects in the Universe (see Frank
et al. 2002 for a recent review). But the luminosity of a nonexplosive
source cannot exceed by large factors the Eddington value $L_{\rm
Edd}$ at which its radiation would drive away the accreting matter
powering it (see Shaviv, 1998, 2000, and Begelman, 2002 for proposals for
exceeding $L_{\rm Edd}$ by factors $\sim 1 - 10$).

Recent observations of external galaxies have revealed ultraluminous
X--ray sources (ULXs: see Makishima et al., 2000 and references
therein) with apparent luminosities well in excess of the Eddington
limit $L_{\rm Edd}$ for a stellar--mass black hole (or neutron
star). This has prompted suggestions of a population of black holes
with masses intermediate between stellar values and the $> 10^6\msun$
inferred for active galactic nuclei (e.g. Ebisuzaki et al., 2001;
Miller \& Hamilton, 2002; Schneider et al., 2001).  Although such a
population is attractive for some pictures of galaxy formation
(e.g. Madau \& Rees, 2001) it is difficult both to create and fuel
enough accreting versions to account for the numbers now observed
(King et al., 2001). In particular the 7 ULXs found in {\it Chandra}
observations of the Antennae (Fabbiano et al., 2001) strongly indicate
a connection with recent massive star formation. In the
intermediate--mass black hole picture it is hard to see why such a
connection should exist (and not be present when massive star
formation has decreased) unless the intermediate--mass black holes are
accreting from massive stars. Neither direct formation nor capture (if
for example the intermediate--mass black holes are the endpoints of
massive Pop III stars) seems a plausible route for making such
binaries (see the discussion in King et al., 2001). In addition, King
\& Davies (2002) have shown that gravitational radiation reaction
makes it extremely difficult to build up such black hole masses by
mergers in globular clusters.

In line with this, an alternative view (King et al., 2001) holds that
ULXs represent a bright, shortlived, but common phase of stellar--mass
X--ray binary evolution. Their high apparent luminosities may either
result from viewing a significantly anisotropic radiation pattern at a
favourable angle, or be genuinely super--Eddington (cf Shaviv 1998,
2000; Begelman, 2002) or both. This picture is consistent with many
observed ULX properties such as their inferred birthrates, spectra and
optical counterparts (Roberts et al., 2001), and even possible
periodicities (Sugiho et al., 2001).

In their original paper King et al. (2001) suggested that the ULX
population might correspond to the phase of thermal--timescale mass
transfer inevitable in high--mass X--ray binaries. The wealth of data
in {\it Chandra} observations now makes it clear that this can only be
a part of the story. In particular sources with luminosities $\ga
L_{\rm Edd}$ for a $20\msun$ black hole have been observed in
elliptical galaxies, which do not contain high--mass X--ray binaries
(e.g. Sarazin, Irwin \& Bregman, 2001; Finoguenov \& Jones, 2001;
Colbert \& Ptak, 2002).

The question of identifying the ULXs is particularly acute if their
radiation patterns are genuinely anisotropic, as this clearly leaves
open the question of how these objects appear from a less special
viewpoint. The possibility of genuine super--Eddington emission does
not remove this problem, as one might expect such sources to radiate
with significantly anisotropy. There is direct support for this view
from observations of SS433. Everything suggests that the mass transfer
rate here is well above the value ($10^{-7}\msun {\rm yr}^{-1}$)
which would power the Eddington limit for a $10\msun$ black hole (cf
King, Taam \& Begelman, 2000) yet the system shows no direct evidence
of exceeding $L_{\rm Edd}$. Thus {\it if} the luminosity of SS433 does
exceed $L_{\rm edd}$ (rather than for example blowing away all of the
excess accreting gas), it clearly does so in directions inaccessible
to us. 

Here I suggest that ULXs are simply X--ray binaries with
super--Eddington accretion rates, and that they can be identified with
two known types of accreting sources: luminous persistent systems such
as SS433, and microquasars (Mirabel \& Rodr\'iguez, 1999) with very
bright prolonged outbursts, such as GRS~1915+105. The first class is
likely to predominate in galaxies with young stellar populations, and
the second in ellipticals. If the accretion processes powering the
brightest black holes are essentially identical whatever the mass
scale, there may exist an apparently ultraluminous class of AGN
closely analogous to the ULXs -- possibly the ultrasoft AGN -- and a
class of Doppler--boosted ULXs.

\section{The Cause of ULX Behaviour}

To identify the parent systems of stellar--mass ULXs we need some idea
of what produces their defining feature, the apparent (or real)
super--Eddington luminosity. A suggested accretion flow producing this
(Jaroszynski et al., 1980, Abramowicz et al., 1980) and invoked in the
original paper (King et al., 2001) on stellar--mass ULXs postulates an
accretion disc whose inner regions are geometrically thick, and a
central pair of scattering funnels through which the accretion
radiation emerges. Note that this form of `beaming' does not involve
relativistic effects, although Doppler boosting in a relativistic jet
has also been suggested as a way of explaining the high luminosities
(Koerding et al., 2001, Markoff et al., 2001). The thick--disc plus
funnels anisotropy mechanism explicitly requires a high mass inflow
rate near the black hole or neutron star accretor, much of which must
be ejected, probably some of it in the form of a jet. (In fact the
motivation of the original papers (Jaroszynski et al., 1980,
Abramowicz et al., 1980) was to produce a geometry favouring jet
production.) Super--Eddington mass inflow is of course trivially
required if the ULX luminosity genuinely exceeds $L_{\rm Edd}$. The
Doppler--boosting anisotropy mechanism does not necessarily require
high mass inflow, and indeed was explicitly used (Markoff et al.,
2001) to model a significantly sub--Eddington outburst in the soft
X--ray transient (SXT) XTE~J1118+480. Nonetheless it is clear that a
high mass inflow rate will lead to higher apparent luminosities in
this case too.

Accordingly I shall henceforth assume that the basic reason for ULX
behaviour is a highly super--Eddington mass inflow rate near the
accretor, leading to three characteristic features: (i) the total
accretion luminosity is of order $L_{\rm Edd}$, (ii) this is confined
to a solid angle $4\pi b \la 4\pi$, making the source apparently
super--Eddington when viewed from within this solid angle (even if it
is not genuinely super--Eddington), and (iii) the bulk of the
super--Eddington mass inflow is either accreted at low radiative
efficiency, or more probably, ejected in the form of a dense outflow,
probably including relativistic jets.

\section{ULX Parent Systems}

The identification with super--Eddington mass inflow rates made above
allows us to identify the likely ULX parent systems. There are two
situations in which X--ray binaries naturally have such rates: phases
of thermal--time mass transfer, and bright SXT outbursts. (I am
indebted to Dr G.A. Wynn for pointing out the relevance of this second
possibility.) The first of these is considered extensively by King et
al. (2001) and its main features can be summarized briefly here.

\subsection{Thermal--timescale mass transfer}

Thermal--timescale mass transfer occurs in any Roche--lobe--filling
binary where the ratio $q$ of donor mass to accretor mass exceeds a
critical value $q_{\rm crit} \sim 1$. Thus all high--mass X--ray
binaries will enter this phase once the companion fills its Roche
lobe, either by evolutionary expansion, or by orbital shrinkage via
angular momentum loss. Depending on the mass and structure of the
donor, extremely high mass transfer rates $\dot M_{\rm tr} \sim
10^{-7} - 10^{-3}\msun {\rm yr}^{-1}$ ensue. SS433 is an example of a
system currently in a thermal--timescale mass transfer phase (King et
al., 2000) which has descended by this route. The idea that SS433
itself might be a ULX viewed `from the side' provides a natural
explanation of its otherwise puzzlingly feeble X--ray emission
($L_{\rm x} \sim 10^{36}$~erg~s$^{-1}$, Watson et al., 1986).

The binary probably survives the thermal--timescale phase without
entering common--envelope (CE) evolution provided that the donor's
envelope is largely radiative (King \& Begelman, 1999). Observational
proof of this is provided by Cygnus X--2 (King \& Ritter, 1999;
Podsiadlowski \& Rappaport, 2000), whose progenitor must have been an
intermediate--mass binary (companion mass $\sim 3\msun$, neutron star
mass $\sim 1.4\msun$). CE evolution would instead have engulfed the
binary and extinguished it as a high--energy source. The binary would
probably have merged, producing a Thorne--Zytkow object.

The birthrates of intermediate and high--mass X--ray binaries are
compatible with the observed numbers of ULXs: King et al. (2001) show
that the birthrates required to explain the latter are independent of
the dimensionless beaming and duty--cycle factors $b, d$. For massive
systems the thermal timescale lasts longer than the preceding
wind--fed X--ray binary phase; the fact that there are far fewer
observed ULXs than massive X--ray binaries must mean that the beaming
and duty--cycle factors obey $bd << 1$.

\subsection{Bright SXT outbursts}

While thermal--timescale mass transfer probably accounts for a
significant fraction of observed ULXs, bright SXT outbursts will also
produce super--Eddington accretion rates. The outbursts themselves
result from the usual hydrogen--ionization disc instability, greatly
prolonged by the effect of irradiation by the central X--rays from the
neutron--star or (more commonly) black--hole accretor. These prevent
the launching of the cooling wave which turns off the instability in
white dwarf systems. Trapped in the hot state, the disc matter must
continue to flow inwards until the central disc is severely depleted,
on a viscous timescale. This reduces the irradiation effect and
eventually allows a return to quiescence.

The condition for SXT behaviour is simply the presence of ionization
zones. This amounts to asking if in a steady state the disc edges are
cool, even when the potentially stabilizing effect of irradiation is
allowed for. King, Kolb \& Burderi (1996) and King, Kolb \&
Szuszkiewicz (1997) show that most low--mass X--ray binaries (LMXBs)
will be transient: in particular this is inevitable for any LMXB with
orbital period $P \ga 2$~d. During an outburst the central inflow rate
is initially of order $\dot M_c \sim \Delta M/t_{\rm visc}$, where
$\Delta M$ is the heated disc mass and $t_{\rm visc}$ its viscous
timescale, and decays exponentially on $t_{\rm visc}$ (King \& Ritter,
1998; Dubus et al., 2001). King \& Ritter (1998, eq 29) show that
$\dot M_c$ will exceed the value $\sim 10^{-7}\ \msun{\rm yr}^{-1}$
powering the Eddington limit for a $10\msun$ black hole if the heated
disc region is larger than about $10^{11}$~cm, corresponding to an
orbital period $P \ga 1$~d. 

SXT outbursts in systems with such periods are complex because of the
large reservoir of unheated mass at the edge of the disc, which can
eventually contribute to the outburst. Full numerical calculations
will be needed to describe this process. However the trends with
increasing $P$ are clear: the outbursts become longer (several decades)
and involve more mass, but the quiescent intervals increase more
rapidly (several $\ga 10^3$~yr) so that the outburst duty cycle $d$
decreases (Ritter \& King, 2001). This results in inflow rates which
become ever more super--Eddington at large $P$. If the accretor is a
neutron star, the resulting inefficient accretion makes it difficult
to spin it up; this probably accounts for the absence of millisecond
pulsars in wide binaries with circular orbits and $P \ga 200$~d
(Ritter \& King, 2001).

Spectacular evidence of super--Eddington accretion is provided by
GRS~1915+105, which has been in effectively continuous outburst since
1992. The observed X--ray luminosity $L_{\rm x} \ga 1\times
10^{39}$~erg~s$^{-1}$ implies that at least $\sim 10^{-6}\msun$ has
been accreted over this time, requiring a large and massive accretion
disc. In line with this, it appears that the binary is wide ($P \simeq
33$~d; Greiner, Cuby \& McCaughrean, 2001). Even at the reported
accretor mass $M_1 = (14 \pm 4)\msun$ (Greiner et al., 2001) there is
little doubt that the current mass inflow near the black hole is
highly super--Eddington.  Evolutionary expansion of the donor will
drive a persistent mass transfer rate $-\dot M_2 \sim 10^{-9}(P/{\rm
d})\msun{\rm yr}^{-1} \sim 3 \times 10^{-8}\msun{\rm yr}^{-1}$ (King,
Kolb \& Burderi, 1996) which is already close to the Eddington rate
$\dot M_{\rm Edd} \sim 10^{-7}\ \msun{\rm yr}^{-1}$. Given an outburst
duty cycle $d << 1$, the mean inflow rate $\sim -\dot M_2/d$ is
$>>\dot M_{\rm Edd}$. Note that we definitely do not look down the jet
in GRS~1915+105, which is at about $70^{\circ}$ to the line of sight
(Mirabel \& Rodr\'iguez, 1999), so it is quite possible that the
apparent luminosity in such directions is much higher than the
observed $L_{\rm x}$. 

Assuming that most of the LMXBs in elliptical galaxies are transients
observed in long--lasting outbursts (Piro \& Bildsten, 2002), this
class of ULXs must obey $b << 1$, since most of these LMXBs are not
ULXs.

\section{Discussion}

I have suggested that the basic cause of ULX behaviour is a highly
super--Eddington mass inflow rate near the accreting component of an
X--ray binary. This occurs in two types of systems, exemplified by
SS433 and microquasars like GRS~1915+105 respectively. The observed
ULX population is a varying mixture of the two, depending on the star
formation history of the host galaxies. SS433--like systems should
predominate in galaxies with vigorous star formation, such as the
Antennae, while ULXs in elliptical galaxies must be of the microquasar
type, as these galaxies contain no high--mass X--ray binaries. We
therefore expect ULXs in ellipticals to be variable. However the
microquasar systems most likely to be identified as ULXs are clearly
those with the brightest and longest outbursts, so baselines of
decades may be needed to see significant numbers turning on or
off. There is some evidence of such variability from the differences
between ROSAT and {\it Chandra} observations of the same galaxies.
The fact that none of the SXTs found in the Galaxy has turned out to
be a ULX suggests that the beaming factor $b$ must be $\la 0.1$ for
this mode of accretion. This agrees with our conclusion above that $b
<< 1$ for the ULXs in ellipticals.

Evidence that the two suggested classes of ULXs do resemble each other
in similarly super--Eddington accretion states comes from Revnivtsen
et al. (2002), who report RXTE observations of an episode of
apparently super--Eddington accretion in the soft X--ray transient
V4641 Sgr. Revnivtsen et al. remark on the similarity of the object's
appearance to SS433 in this phase. One might be discouraged by the
apparent suppression of X--rays in this state. However we are
presumably outside the beam of most intense X--ray emission in both
cases: neither should actually appear as a ULX. More work is needed on
whether the X--ray spectra from these objects are consistent with
X--rays leaking sideways from the assumed accretion geometry. Direct
evidence that X--ray emission in ULXs is anisotropic is perhaps
understandably meagre, but may be suggested by the comparison of
optical and X--ray data in NGC 5204 X--1 (Roberts et al., 2002), where
low--excitation optical spectra are seen from regions close to the
ULX.

Both SS433 and the microquasars are distinguished by the presence of
jets, at least at some epochs. In SS433 the jets precess with a
164--day period, presumably because of disc warping (Pringle,
1996). If looking closely down the jet is required in order to see
high luminosities one might expect to see such periods in a class of
ULXs. This effect could for example explain the $\sim 106$~d
modulation seen in the bright source in M33 (M33 X--8, $L_{\rm x} \sim
10^{39}$erg~s$^{-1}$) by Dubus et al. (1999). However a beam as narrow
as commonly inferred ($\la 1^{\circ}$) for SS433 would give an
unacceptably short duty cycle. If instead it is not necessary to look
down the jet to see a high luminosity, this would rule out Doppler
boosting as the cause of the latter, and ULXs would not be direct
analogues of BL~Lac systems.

This tentative conclusion suggests another. X--ray binaries and active
galactic nuclei share the same basic model, and so far have shown a
fairly good correspondence in their modes of behaviour. If as
suggested above ULXs do not correspond to BL~Lac systems, this may
mean that we are currently missing a class of each type: there should
exist an apparently super--Eddington class of AGN, and a set of X--ray
binaries with Doppler--boosted X--ray emission. King \& Puchnarewicz
(2002) have suggested that the ultrasoft AGN may have apparently
super--Eddington luminosities. A search for for Doppler--boosted
X--ray binaries among the ULXs may be rewarding.

\section{Acknowledgments}

I thank members of the theoretical astrophysics group, particularly
Graham Wynn, for helpful discussions. Research in theoretical
astrophysics at Leicester is supported by a PPARC rolling grant.

\end{document}